\documentstyle[preprint,aps]{revtex}
\preprint {IMSc-94/24}
\begin{document}
\draft
\title{
Thermodynamics of an one-dimensional ideal gas with fractional exclusion
statistics
}

\author{ M. V. N. Murthy
 and R. Shankar}

\address
{The Institute of Mathematical Sciences, Madras 600 113, India.\\
}
\date{\today}
\maketitle
\begin{abstract}
We show that the particles in the Calogero-Sutherland Model obey fractional
exclusion statistics as defined by Haldane.  We construct anyon number
densities and derive the energy distribution function. We show that the
partition function factorizes  in the form characteristic of an ideal gas.
The virial expansion is exactly computable  and interestingly it is only the
second virial coefficient that encodes the statistics information.

\end{abstract}

\pacs{PACS numbers: 05.30.-d, 71.10.+x}

\narrowtext
There has been some progress recently \cite{ms,wu} in understanding
systems of particles with fractional exclusion statistics as defined
by Haldane\cite{haldane}.  Motivated by  physical examples, such as
quasi-particles in the fractional quantum Hall systems and spinons in
antiferromagnetic spin chains, Haldane formulated a generalized
exclusion principle as follows: He considered  many particle systems
with finite dimensional Hilbert spaces where the dimension of the
single particle Hilbert space
depends linearly on the total number of particles present. Then the
exclusion statistics parameter  $g$  is defined by $\Delta d = -g
\Delta N$, where $\Delta d$ is the change in the dimension of the
single particle space and $\Delta N$ is the change in the number of
particles. Thus  $g$  is a measure of the (partial) Pauli blocking in
the system - $g=0(1)$ corresponds to bosons (fermions). In a recent
work\cite{ms} we have  generalized the definition of  $g$  to the case of
infinite dimensional Hilbert spaces so as to apply Haldane's
definition to systems in the continuum with no sharp cutoff in energy.
If for such a system a virial expansion exists and all the virial
coefficients are finite in the high temperature limit, then the
exclusion statistics parameter  $g$
is completely determined by the second
virial coefficient in this limit.  In many systems it is  also possible to
relate  $g$  to the exchange statistics parameter $\alpha$.

In a remarkable paper\cite{wu}, Wu has recently considered a system
with flat dispersion and assuming that Haldane exclusion principle
holds (as in the case of anyons in a magnetic field in two dimensions
confined to the lowest Landau level) has shown that the statistical
distribution function  of this system is,
\begin{equation}
\bar n(\epsilon) = \frac{1}{w(\epsilon) +g} .
\label{dist}
\end{equation}
Here $\epsilon$ is the single particle energy,
and $w(\epsilon)$ is given as the solution of the
equation,
\begin{equation}
w^g(\epsilon,g)(1+w(\epsilon,g))^{1-g} = e^{\beta(\epsilon-\mu)},
\label{wofg}
\end{equation}
where $\beta$ is the inverse temperature and $\mu$ is the chemical
potential of the system.
The distribution function smoothly interpolates between the  Bose ($g=0$)
and Fermi($g=1$) distribution functions.  It therefore seems likely
that the distribution function discovered by Wu may be valid in a more
general context. Indeed it could be the general form for
the ideal (exclusion) anyon\cite{note1} distribution function. We will
demonstrate that this is so for a model hamiltonian system in one
space dimension with inverse square law potential- the
Calogero-Sutherland Model(CSM)\cite{csm}.

It has lately been recognized that models with the inverse square law
interaction, the CSM\cite{csm}, Haldane-Shastry\cite{hs} model and
other related models behave very much like ideal gases and that
the particles have fractional statistics\cite{hahata}.  In this paper,
we concentrate on the CSM, which is a model defined in the continuum.
Though the spectrum of the model and its thermodynamics are exactly
known\cite{csm} the ideal anyon gas nature has not been brought out so
far.  To this end we apply our definition of exclusion
statistics\cite{ms} to it and show that the
particles have fractional exclusion statistics. Next we define the
anyon number density in the energy space and prove that, in the
thermodynamic limit, its average is given by the distribution of the
form discovered by Wu.  We then show that the partition function
factorizes into factors obtained by integrating the
distribution function with respect to the single particle energies.
This clearly demonstrates the ideal gas nature of the system. Finally,
we comment on an interesting feature of the exact equation of the
state.

We work in the fermionic basis of the CSM.  The hamiltonian of the
system of interacting fermions in this model is given by ($\hbar =1$),
\begin{equation}
H = \sum_{i=1}^N \left [ -\frac{1}{2} \frac{\partial^2}{\partial x_i^2} +
\frac{1}{2} \omega^2 x_i^2 \right ] +
\frac{1}{2} \sum_{i<j=1}^N \frac{(g^2-1)}{(x_i-x_j)^2},
\label{ham}
\end{equation}
where the particles are confined in a harmonic well and the
thermodynamic limit is obtaining by taking $\omega \rightarrow 0$.
The particles can also be put on a circle with $|x_i-x_j|$ replaced by
the chord length $|sin \frac{\pi}{L}(x_i-x_j)|$ in the interaction
term.  This model has the same thermodynamic limit as the one in
eq.(\ref{ham}). At special values of the coupling the model can be mapped on to
particular matrix models\cite{sa}. It has recently been shown that a system of
2-d anyons
(defined through the exchange phase) confined to the rim of a disc can
also be mapped on to the model on a circle\cite{li}.

The spectrum of CSM is exactly known.  The states can be labelled by a
set of fermionic occupation numbers $\{n_k\},~ k=1,...,\infty,~ n_k =
0,1$.  The energy is given by,
\begin{equation}
E[\{n_k\}] = \sum_{k=1}^{\infty} \epsilon_k n_k - \omega  (1-g)
\frac{N(N-1)}{2},
\label{energy}
\end{equation}
where $\epsilon_k = k\omega,~ N=\sum_{k=1}^\infty n_k$.  We note that
this spectrum is identical to the spectrum of quasi-particle states in
a gaussian theory of compact bosons\cite{ms}  with radius $R =
1/\sqrt(g)$  with the following identification- while $k$
in the above equation is a state label in CSM, in the gaussian theory
of compact bosons $k$ refers to the box quantized momenta.  The
following discussions thus applies to this case also.  As can be seen
from eq.(\ref{energy}), the effect of the interaction is that each
particle shifts the energy of every other particle by a constant
$\omega (g-1)$.  The energy functional can also be written as,
\begin{equation}
E[\{n_k\}] = \sum_{k=1}^{\infty} \epsilon_k n_k - \omega  (1-g)
\sum_{k_1<k_2=1}^{\infty} n_{k_1} n_{k_2}.
\label{enfun}
\end{equation}
The exact spectrum of the model is thus reproduced by an effective
hamiltonian of quasi-particle with constant density of states and
constant Landau parameters\cite{gb}.  As we had discussed in our earlier paper
\cite{ms} this scale invariant energy shift is the basic reason for
the occurrance of nontrivial exclusion statistics.  We had also shown
that a spectrum of the form in eq.(\ref{energy}) results in the
exclusion statistics parameter  being equal to  $g$ .

Thus the particles in the CSM are anyons in the sense of Haldane with
the exclusion
statistics parameter equal to g.  We will now  define anyon number
densities
and show that their thermal average  is exactly given by the
distribution function
in eq.(\ref{dist}).   To this end, we define $N(\epsilon,0)$ as the number of
particles with energy $\epsilon_k < \epsilon$, ie., $N(\epsilon,0) =
\sum_{k=1}^{\infty} \theta (\epsilon - \epsilon_k)n_k$, where $\theta
(x)=0$ for $x \le 0$ and 1 for $x>0$.  We may now define the shifted
single particle energies as
\begin{equation}
\epsilon_A(k,g) = \epsilon_k - \omega (1-g)N(\epsilon_k,0).
\label{enany}
\end{equation}
We identify this shifted energy with anyon single particle energy since
the total energy can now be written as
\begin{equation}
E[\{n_k\}] = \sum_{k=1}^{\infty} \epsilon_A(k,g) n_k
\label{entot}
\end{equation}
The number of particles with energy less than $\epsilon$ is then
 $N(\epsilon,g) =
\sum_{k=1}^{\infty} \theta (\epsilon - \epsilon_A(k,g))n_k$.  The
anyon number density in the thermodynamic limit is then given by
\begin{equation}
n_A(\epsilon,g) = \lim_{\Delta \epsilon \rightarrow 0}\lim_{\omega
\rightarrow 0}
\frac{N(\epsilon+\Delta \epsilon, g)-N(\epsilon ,g)}{\Delta \epsilon},
\label{ndef}
\end{equation}
where the limit $\omega \rightarrow 0$ is to be taken first.

We will now derive a differential equation for $\bar N(\epsilon,g)$ and
solve it to obtain the anyon distribution function.  Consider a state
with N-particles labelled by $\{k_i\}, i=1,...,N$ ordered such that
$k_{i+1} > k_i$.  Thus for this state $N(\epsilon_{k_i},0) = i-1$.
The $i^{th}$ shifted energy is  $\epsilon_A(k_i,g) = \epsilon_{k_i} - \omega
(1-g)(i-1)$. Note that $\epsilon_A(k_{i+1},g) - \epsilon_A(k_i,g) \ge
\omega g > 0$.
Thus the anyon energies $\epsilon_A(k_i,g)$  also increase monotonically
with i. We therefore have,
\begin{equation}
N(\epsilon_{k_i},g) = N(k_i,0) = i-1
\label{Ns}
\end{equation}
To get some feel for the anyon number densities, let us consider the
ground state of the N particle system. The set $\{k_i\}$ is (1,2,3,...,N).
The corresponding set of shifted energies is
$(\omega,(1+g)\omega,(1+2g)\omega,...,(1+(N-1)g)\omega)$. The density of
particles at energy $\epsilon$ is then,
\begin{eqnarray}
n_A(\epsilon,g) & = & {1 \over g}~~,~~~~  \epsilon < \epsilon_F = g\bar{\rho}
\nonumber \\
                & = & 0~~,~~~~    \epsilon \geq \epsilon_F
\label{na0temp}
\end{eqnarray}
Where $\bar{\rho} \equiv \omega N$ is the average density. This is the
same as the zero temperature limit of the distribution function in
eq.(\ref{dist}).

Next we compute how the anyon density changes with $g$. From eqs.(\ref{ndef})
and (\ref{Ns}) we see that as $g$ increases the particles move
to the right in $\epsilon$ space with
a `velocity' given by $\omega N(\epsilon,g)$.  Thus the number of
particles crossing a point $\epsilon$ when  $g$  increases by $\Delta g
$, is given by the velocity at the point of crossing multiplied by the
density at that point.  We thus obtain the differential equation,
\begin{equation}
\frac{\partial \rho (\epsilon,g)}{\partial g } = \rho
(\epsilon,g)\frac{\partial \rho (\epsilon,g)}{\partial \epsilon } ,
\label{diffeqn}
\end{equation}
where $\rho(\epsilon,g) = \omega N(\epsilon,g)$.  Denoting the thermal
average of $\rho(\epsilon,g)$ as $\bar\rho(\epsilon,g)$ and using
the fact that in the thermodynamic limit, $(\bar \rho^2(\epsilon,g)) =
(\bar \rho (\epsilon,g))^2$ we obtain the differential equation,
\begin{equation}
\frac{\partial \bar \rho (\epsilon,g)}{\partial g } = \bar \rho
(\epsilon,g)\frac{\partial \bar \rho (\epsilon,g)}{\partial \epsilon }.
\label{diffeqnave}
\end{equation}
The distribution function $\bar n_A(\epsilon,g)$ can be obtained from
the solution of eq.(\ref{diffeqnave}) by using $\bar n_A(\epsilon,g) =
\frac{\partial \bar \rho (\epsilon,g)}{\partial \epsilon}$.  Thus we
need the solution to eq.(\ref{diffeqnave}), with the boundary condition,
\begin{equation}
\frac{\partial \bar \rho (\epsilon,g)}{\partial \epsilon}
\rule[-0.4cm]{0.15mm}{1cm}_{g=1} =
\frac{1}{e^{\beta \epsilon} + 1}.
\label{bcondn}
\end{equation}
We will now show that eq.(\ref{diffeqnave}) along with the boundary
condition (\ref{bcondn}) is satisfied by ,
\begin{equation}
\bar \rho (\epsilon,g) = \int_0^{\epsilon} d\epsilon'
\frac{1}{w(\epsilon',g) + g}~~,
\label{soln}
\end{equation}
where $w(\epsilon,g)$ is determined through eq.(\ref{wofg}).
By changing variables from $\epsilon$ to $w$ in eq.(\ref{soln}) the
integral can be done to get,
\begin{equation}
\bar{\rho}(\epsilon,g) = {1 \over \beta}(ln({w(\epsilon) \over
1+w(\epsilon)}) - ln({w(0) \over 1+w(0)}))
\label{rhoint}
\end{equation}
The condition $\bar{\rho}(\infty,g) = \bar{\rho}$ and the fact that
$\lim_{\epsilon \rightarrow \infty} w(\epsilon,g)\rightarrow \infty$
implies that
$w(0)$ is independent of $g$ and $e^{-\beta \mu} = w(0)^g(1+w(0))^{1-g}$. It
can be verified that,
\begin{equation}
{\partial \over \partial g}w(\epsilon) = {(w+g) \over
w(1+w)}\bar{\rho}(\epsilon,g),
\label{dwg}
\end{equation}
\begin{equation}
{\partial \over \partial \epsilon}w(\epsilon) = {1 \over \beta} {(w+g)
\over w(1+w)}.
\label{dwep}
\end{equation}
Using the above results it can be verified that  the form of
$\bar{\rho}(\epsilon ,g)$ in eq.(\ref{soln}) does satisfy
eq.(\ref{diffeqnave}).
Thus the anyon distribution
function in the CSM is given by,
\begin{equation}
\bar n_A(\epsilon,g) = \frac{1}{w(\epsilon,g)+g}
\label{newdist}
\end{equation}
which is exactly the distribution function derived by Wu \cite{wu}
for the case of constant spectral distribution.  The additivity
property of the anyon energies suggests that the
grand partition function should be expressible in a factorized form.
If we assume that the usual definition of the distribution function for the
average number of particles holds for anyons, namely,
\begin{equation}
\bar n_A (\epsilon,g) = {\partial \over \partial \epsilon} ln z(\epsilon)
\end{equation}
where $ln Z^A = \int_0^{\infty} d\epsilon~ln z(\epsilon)$. Then it
follows that we must
have,
\begin{equation}
ln Z^A = \int_0^\infty d\epsilon~~ ln (1+ w^{-1} (\epsilon))
\label{gpfa}
\end{equation}
If this were true, it would show that the particles in CSM have the
basic properties of an ideal gas.  We now show that this is indeed the
case.

The N-particle partition function for the spectrum in eq.(\ref{energy})
is given by
\begin{equation}
Z_N^A = e^{\tilde \beta (1-g) \frac{N(N-1)}{2}} Z_N^F,
\label{zna}
\end{equation}
where $\tilde \beta = \beta \omega$ and $Z_N^F$ is the N particle
fermion partition function.  Setting $g=0$, the bosonic partition
function is obtained,
\begin{equation}
Z_N^B = e^{\tilde \beta \frac{N(N-1)}{2}} Z_N^F,
\label{znb}
\end{equation}
Combining eqs.(\ref{zna}) and (\ref{znb}) we may write the anyon partition
function as,
\begin{equation}
Z_N^A = (Z_N^F)^g (Z_N^B)^{1-g}.
\label{znag}
\end{equation}
The grandpartition function may also be written in the form,
\begin{equation}
Z^A = \sum_{N=0}^{\infty} e^{\beta \mu_A N} Z_N^A =
\sum_{N=0}^{\infty}(e^{\beta \mu_F N}(Z_N^F))^g (e^{\beta \mu_B
N}(Z_N^B))^{1-g},
\label{gznag}
\end{equation}
where we have set the anyon chemical potential $\mu_A = g\mu_F
+(1-g)\mu_B$.  In the thermodynamic limit, the sum is
saturated at the value of $N = \bar {N}$ where $\bar{N}= {1 \over
\beta} {\partial ln
Z^A \over \partial \mu_A}$. The grand partition function can then be
written as,
\begin{equation}
Z^A(\beta,\mu) = (Z^F(\beta,\mu_F))^g(Z^B(\beta,\mu_B))^{1-g},
\label{Zprod}
\end{equation}
where $\mu_F$ and $\mu_B$ are determined by the conditions, $\bar{N}={1 \over
\beta}{\partial ln~Z^F \over \partial \mu_F} = {1 \over
\beta}{\partial ln~Z^B \over
\partial \mu_B}$. The solution to these conditions is $e^{-\beta\mu_F} = w(0),
e^{-\beta\mu_B} = 1+w(0)$. It can be verified that this ensures that ${1 \over
\beta}{\partial
lnZ^A \over \partial \mu_A} = \bar{N}$. We therefore have
\begin{equation}
ln~Z^A(\beta,\mu) = g\int_0^{\infty} dk~ln(1+e^{\beta (k-\mu_F)}) -
(1-g)\int_0^{\infty} dk~
ln(1-e^{\beta (k-\mu_B)})
\label{lnZprod}
\end{equation}
We now make the change of variables $(1+e^{\beta (k-\mu_F)}) = 1+
w^{-1}$ in the
first term and $(1-e^{-\beta (k-\mu_B)}) = 1- (1+w)^{-1}$ in the second term.
We then
have,
\begin{equation}
ln~Z^A(\beta,\mu) = \int_{w(0)}^{\infty} dw {w+g \over w(1+w)} ln(1+w^{-1})
\label{lnZdw}
\end{equation}
We now again make a change of variable from $w$ to $\epsilon$ using
eq.(\ref{wofg}).
The partition function then gets written as,
\begin{equation}
lnZ^A = \int_0^{\infty}d\epsilon~ ln(1+w^{-1}(\epsilon))
\label{lnZfac}
\end{equation}
This is exactly the factorized form in eq.(\ref{gpfa}).

We now consider the equation of state of the ideal gas of anyons in
CSM.  This has already been done by Sutherland\cite{csmt} who however
chose to work with the fugacity expansion. The coefficients of terms in this
expansion in general depend on  $g$  and appear quite complicated.  However,
the
coefficients of the virial expansion are extremely simple. From the
product form of the partition function in eq.  (\ref{Zprod}), the virial
coefficients can all be computed exactly since the virial expansion
for the free
Fermi and Bose gas is exactly computable when the density of states is
constant. We
have,
\begin{equation}
lnZ^A = \beta~ P = \bar{\rho}~ \sum_{l=1}^{\infty} b_l~(\beta
\bar{\rho})^{l-1}
\label{virexp}
\end{equation}
where $b_l$ are the virial coefficients for the ideal Fermi or Bose
gas for $l \neq
2~~ ( ~b_l^F = b_l^B ~~\forall ~~l\neq2$~ for constant density of
states) and
$b_2 = -{1 \over
2}({1 \over 2} - g)$. It is interesting to note that the interaction
affects only the
second virial coefficient. As we had shown \cite{ms}, it is the second
virial
coefficient that determines the exclusion statistics. Thus the ${1
\over r^2}$
interaction in this system modifies the equation of state in a minimal
way. It can
therefore be thought of as a purely statistical interaction in this
system.

In conclusion, we have shown by applying our definition\cite{ms} that
in the CSM, the particles have
fractional exclusion
statistics. We have then constructed the  anyon number
densities whose average is the ideal anyon distribution function
discovered in the
context of systems with a flat dispersion\cite{wu}. We have also shown
that the partition
function can be written in the factorized form characteristic of an ideal gas.
Finally we have computed the virial expansion and have shown
that only the
second virial coefficient is affected by the interaction. The system
thus possesses
all the properties we expect of an ideal gas. Thus we have shown the
system to be an
ideal gas with fractional statistics.


\begin{references}
\bibitem{note1}
Note that we use the name anyon generically for all particles with
fractional statistics- exclusion or exchange.
\end{references}
\end{document}